\documentclass[]{spie}  %>>> use for US letter paper
\usepackage{amsmath,amsfonts,amssymb}
\usepackage{graphicx}
\usepackage[colorlinks=true, allcolors=blue]{hyperref}

\title{CCAT-prime: The Optical Design for the Epoch of Reionization Spectrometer}

\author[a]{Zachary B. Huber}
\author[a,b]{Steve K. Choi}
\author[a]{Cody J. Duell}
\author[b]{Rodrigo G. Freundt}
\author[c]{Patricio A. Gallardo}
\author[a]{Ben Keller}
\author[a,d]{Yaqiong Li}
\author[a]{Lawrence T. Lin}
\author[a,b]{Michael D. Niemack}
\author[e]{Thomas Nikola}
\author[f]{Dominik A. Riechers}
\author[b]{Gordon Stacey}
\author[a]{Eve M. Vavagiakis}
\author[g]{Bugao Zou}
\affil[a]{Department of Physics, Cornell University,  Ithaca, NY 14853, USA}
\affil[b]{Department of Astronomy, Cornell University, Ithaca, NY 14853, USA}
\affil[c]{Kavli Institute for Cosmological Physics, University of Chicago, Chicago, IL 60637 USA}
\affil[d]{Kavli Institute at Cornell for Nanoscale Science, Cornell University, Ithaca, NY 14853, USA}
\affil[e]{Cornell Center for Astrophysics and Planetary Sciences, Ithaca, NY 14853, USA}
\affil[f]{I. Physikalisches Institut, Universit\"at zu K\"oln, Z\"ulpicher Strasse 77, D-50937 K\"oln, Germany}
\affil[g]{Department of Applied and Engineering Physics, Cornell University, Ithaca, NY 14853, USA}

\authorinfo{Further author information: (Send correspondence to Z.B.H.)\\Z.B.H.: E-mail: zbh5@cornell.edu}

\begin{document} 
\maketitle

\begin{abstract}
The Epoch of Reionization Spectrometer (EoR-Spec) will be an instrument module for the Prime-Cam receiver on the CCAT-prime Collaboration’s Fred Young Submillimeter Telescope (FYST), a 6-m primary mirror Crossed Dragone telescope. With its Fabry-Perot interferometer (FPI), EoR-Spec will step through frequencies between 210 and 420 GHz to perform line intensity mapping of the 158 $\mu$m [CII] line in aggregates of star-forming galaxies between redshifts of 3.5 and 8 to trace the evolution of structure in the universe during the epoch of reionization. Here we present the optical design of the module including studies of the optical quality and other key parameters at the image surface. In order to achieve the required resolving power (R$\sim$100) with the FPI, it is important to have a highly collimated beam at the Lyot stop of the system; the optimization process to achieve this goal with four lenses instead of three as used in other Prime-Cam modules is outlined. As part of the optimization, we test the effect of replacing some of the aspheric lenses with biconic lenses in this Crossed Dragone design and find that the biconic lenses tend to improve the image quality across the focal plane of the module.
  
\end{abstract}

% Include a list of keywords after the abstract 

\section{INTRODUCTION}
\label{sec:intro}  % \label{} allows reference to this section

The Epoch of Reionization Spectrometer (EoR-Spec) \cite{Cothard_EoRSpec_2020}\textsuperscript{,} \cite{Nikola_EoRSpec_2022} is an instrument module that will occupy one of the six outer optics tube locations in the Prime-Cam receiver on the CCAT-prime Collaboration’s Fred Young Submillimeter Telescope (FYST) \cite{ccat_science_2021}, a 6-m primary mirror Crossed Dragone telescope. \cite{Niemack_telescope_2016}\textsuperscript{,} \cite{Parshley2018} 
EoR-Spec uses a Fabry-Perot interferometer (FPI) composed of silicon-substrate-based (SSB) mirrors \cite{Bugao_EoRSpecMirrors_2022} located at the Lyot stop of the optical system to perform line intensity mapping of the 158 $\mu$m [CII] line to study the evolution of structure between redshifts 3.5 and 8. These measurements will enable studies of the formation of the first galaxies near the end of the epoch of reionization at higher redshifts and near the peak star-forming epoch at lower redshifts. The FPI steps through frequencies between 210 and 420 GHz and illuminates one broadband, non-polarization sensitive detector array centered around 370 GHz (0.8 mm) and two arrays centered around 260 GHz (1.1 mm). EoR-Spec is slated to begin science operations on FYST in late 2024.

In order to make the best use of this instrument, the design of the cold refracting silicon optical elements that reimage the secondary focus of the telescope onto the focal plane of the instrument module needs to be optimized for high image quality and a well-collimated beam at the location of the FPI. 
Section~\ref{sec:design} provides an overview of the design and section~\ref{sec:performance} quantifies its performance. The use of a biconic lens in this design and its benefits are discussed in section~\ref{sec:biconic} before section~\ref{sec:conclusion} outlines possible avenues for future improvement.

\section{OPTICAL DESIGN}
\label{sec:design}

The optical design\footnote{The optical design and optimization were performed in Zemax OpticsStudio \cite{zemaxwebsite}.} for the EoR-Spec instrument module evolved from the three-lens Simons Observatory (SO) instrument module optical design described in Ref.~\citenum{Dicker2018}, which is identical to the design of the 280 GHz module for Prime-Cam. Like the SO optics design, the goal of the optimization was to produce excellent image quality across the focal plane, a consistent Lyot stop for the system for all fields, a focal plane size near 275 mm to illuminate only the mechanical size of our detector arrays, and angles of incidence on the image plane of less than two degrees to appropriately match the size of our detector array feedhorns. 

In addition to these constraints, this design needs to be diffraction-limited across a wider range of frequencies than the previous design due to the wider range of frequencies covered by the FPI. For this instrument, it is also important to have a highly collimated beam at the Lyot stop of the system to achieve the required resolving power of R$\sim$100 with the FPI. The fabrication of AR coatings for the silicon lenses used in this module is significantly simplified if the sag of the surface of each lens is 14 mm or smaller \cite{Datta_ARcoating_2013}. The final sags for each surface are shown in Table~\ref{sag_table}.

\begin{figure}
    \centering
    \includegraphics[width=0.9\linewidth]{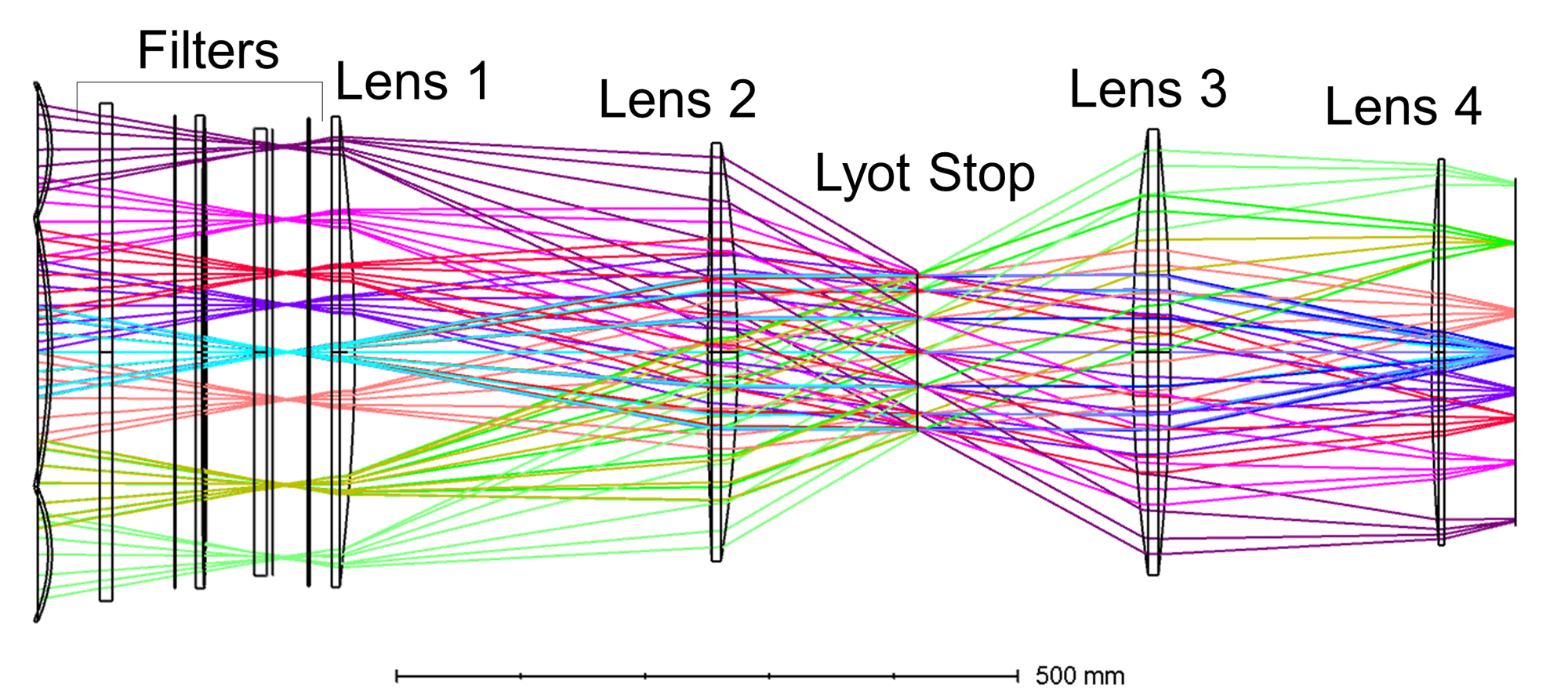}
    \caption{A ray trace of the optics design with selected fields depicted in different colors. The receiver vacuum window is on the far left, while the image plane is on the far right. Each field is highly collimated at the Lyot stop for FPI performance. Lens two and three were switched from planar-convex to convex-convex to make it easier to apply the AR coating.}
    \label{fig:raytrace}
\end{figure}

Moving lens two closer to lens one produced a more highly collimated beam, and the addition of a fourth lens compared to the SO design maintained a high level of image quality across the focal plane for the relevant frequencies. To improve the image quality further, lens three was changed from an aspheric lens to a biconic lens. Lenses two and three were changed to convex-convex to accommodate the sag requirements. Fig.~\ref{fig:raytrace} shows a ray trace of the optics design with these modifications.

\begin{table}
\begin{center}
\begin{tabular}{ |c|c| } 
 \hline
 Surface & Sag (mm)\\ 
 \hline
 Lens 1 & 12.08 \\ 
 \hline
 Lens 2 Front & 3.00 \\ 
 \hline
 Lens 2 Back & 13.35 \\ 
 \hline
 Lens 3 Front & 12.40 \\ 
 \hline
 Lens 3 Back & 9.97 \\ 
 \hline
 Lens 4 & 5.81 \\ 
 \hline
\end{tabular}
\caption{The absolute value of the sags for the different lens surfaces in the design. Lenses 1 and 4 are planar-convex, while lenses 2 and 3 are convex-convex.}
\label{sag_table}
\end{center}
\end{table}

\begin{figure}
    \centering
    \includegraphics[width=0.7\linewidth]{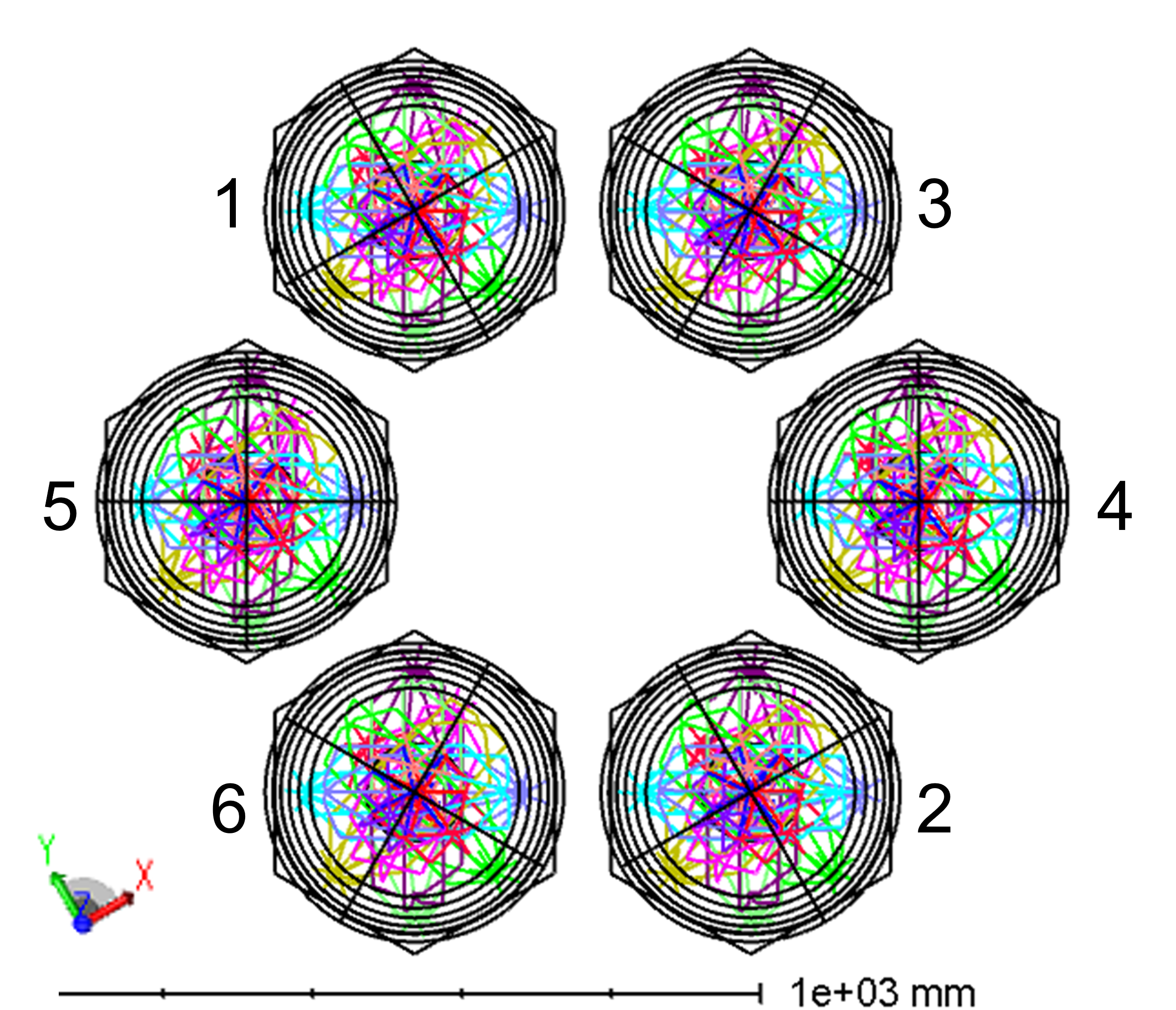}
    \caption{The six outer optics tubes in Prime-Cam where EoR-Spec could be placed, as viewed from the image side (looking out towards the telescope mirrors). The colored lines depict rays from different fields on the sky. Each configuration is labeled with an identifying number and the coordinate system in Zemax for reference to plots below.}
    \label{fig:sixtubesreceiver}
\end{figure}

There are seven possible locations for instrument modules in Prime-Cam. EoR-Spec will be located in one of the outer six instrument module locations. In order to determine which location would have the best optical quality for the needs of this instrument, the optimization simultaneously tests the same optics in six different locations, or configurations, within the receiver, as shown in Fig.~\ref{fig:sixtubesreceiver}. Each tube is rotated to ensure that the x and y axes for the biconic lens always remain in the same orientation with the x axis pointing radially towards the center of the telescope.

In addition to helping us to determine the best location for EoR-Spec in Prime-Cam, testing all six tubes also allows us to simulate the effect of scanning the telescope in elevation. Unlike SO, which uses a co-rotator to keep the receiver in the same orientation relative to the telescope mirrors as the telescope aperture scans in elevation, the effective location of each instrument module in Prime-Cam rotates as the elevation angle changes. The different module locations in the optical design provide information on how the optical properties of each location change as the telescope tilts 60 degrees in elevation. Since 60 degrees is larger than the maximum elevation change that the telescope can scan through during operation, these tubes provide an upper bound on the changes to the optical properties of each optics tube due to the changing elevation of the telescope during observations.

\section{OPTICAL PERFORMANCE OF NOMINAL DESIGN}
\label{sec:performance}
 
\begin{figure}
    \centering
    \includegraphics[width=0.9\linewidth]{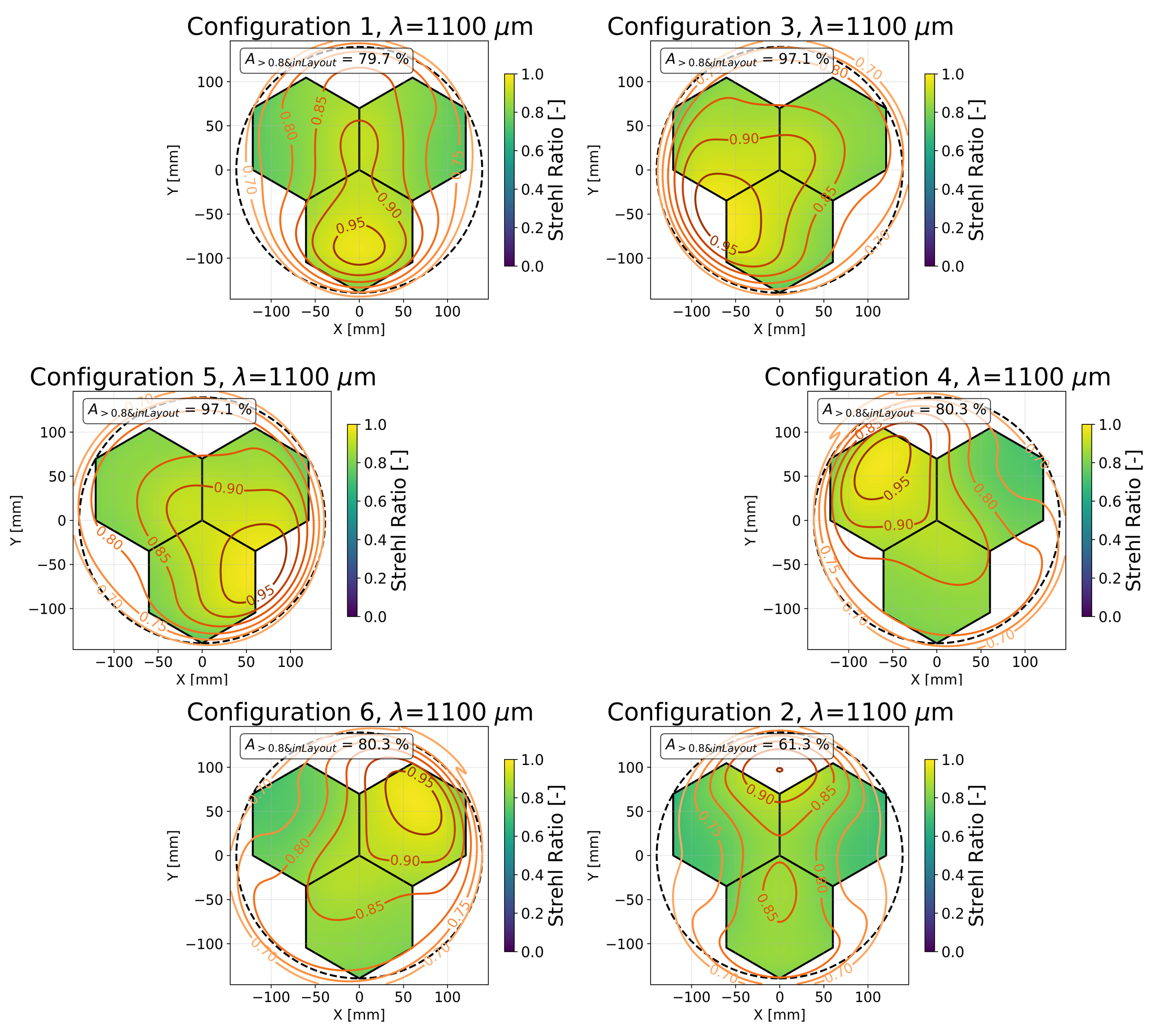}
    \caption{The Strehl ratio across the focal plane at 1.1 mm for each of the six possible locations at an elevation of 60 degrees. Because of the rotation used in the Zemax file for the receiver to simulate the elevation shift away from zenith (which produces the coordinate system shown in Fig.~\ref{fig:sixtubesreceiver}), each plot looks rotated by 30 degrees away from the symmetry axis of the telescope mirrors. The highest Strehl ratio is always in the part of the configuration closest to the center of the receiver. The three hexagons show one possible orientation of the three detector arrays within the instrument module. The percentage of the area of these hexagons that is diffraction-limited (Strehl ratio $>$ 0.8) is reported in the legend, while the contours show increments of 0.05 in Strehl ratio from 0.7 to 1.0. For this wavelength, over 80\% of the detector area is diffraction-limited for most configurations.}
    \label{fig:strehl11mm}
\end{figure}

\begin{figure}
    \centering
    \includegraphics[width=0.9\linewidth]{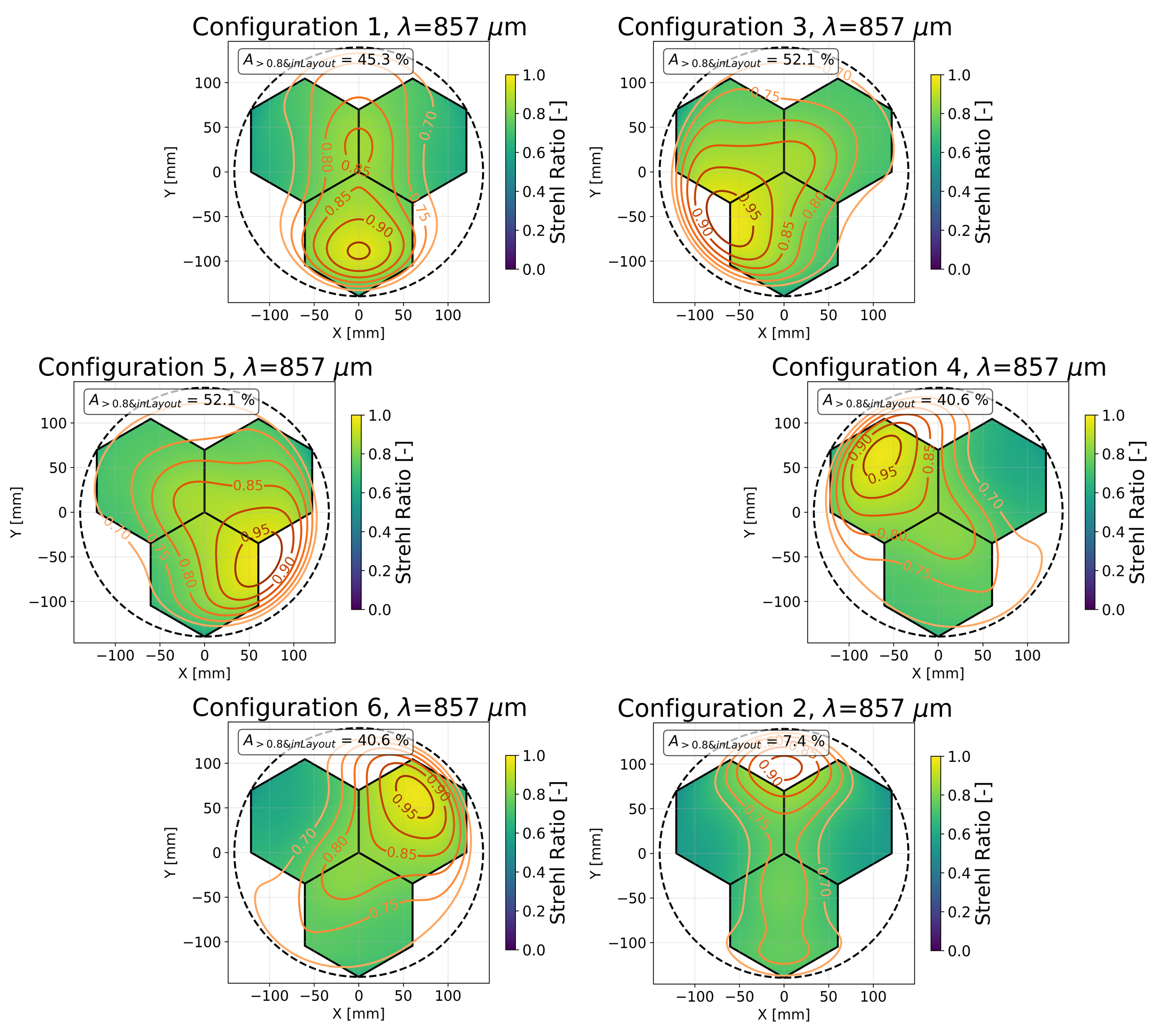}
    \caption{The Strehl ratio across the focal plane at 0.8 mm for each of the six possible locations at an elevation of 60 degrees. Because of the rotation used in the Zemax file for the receiver to simulate the elevation shift away from zenith (which produces the coordinate system shown in Fig.~\ref{fig:sixtubesreceiver}), each plot looks rotated by 30 degrees away from the symmetry axis of the telescope mirrors. The highest Strehl ratio is always in the part of the configuration closest to the center of the receiver. The three hexagons show one possible orientation of the three detector arrays within the instrument module. The percentage of the area of these hexagons that is diffraction-limited (Strehl ratio $>$ 0.8) is reported in the legend, while the contours show increments of 0.05 in Strehl ratio from 0.7 to 1.0. For this wavelength, at least one of the three detector arrays is diffraction-limited in most configurations, meeting our requirements as only one of the three arrays will measure at this higher frequency band.}
    \label{fig:strehl086mm}
\end{figure}

The Strehl ratio, defined as the ratio of the peak intensity of the point spread function in the real system to the peak intensity of the point spread function of the system with aberrations removed \cite{zemaxdocs}, is used to quantify the image quality of the full design, including the telescope mirrors and the cold lenses. The goal of the optimization was to produce a diffraction-limited design, which we define as a Strehl ratio greater than 0.8. As shown in Fig.~\ref{fig:strehl11mm} and Fig.~\ref{fig:strehl086mm} for an elevation of 60 degrees, the design is diffraction-limited for most locations across 80-95\% of the focal plane at 1.1 mm (with much of the remaining area above 0.7) and 30-50\% at 0.8 mm. This meets our requirements for one 0.8 mm detector array and two 1.1 mm detector arrays.
Two locations achieve greater than 90\% coverage at 1.1 mm while simultaneously covering at least one array's worth of focal plane array at 0.8 mm.

\begin{figure}
    \centering
    \includegraphics[width=0.9\linewidth]{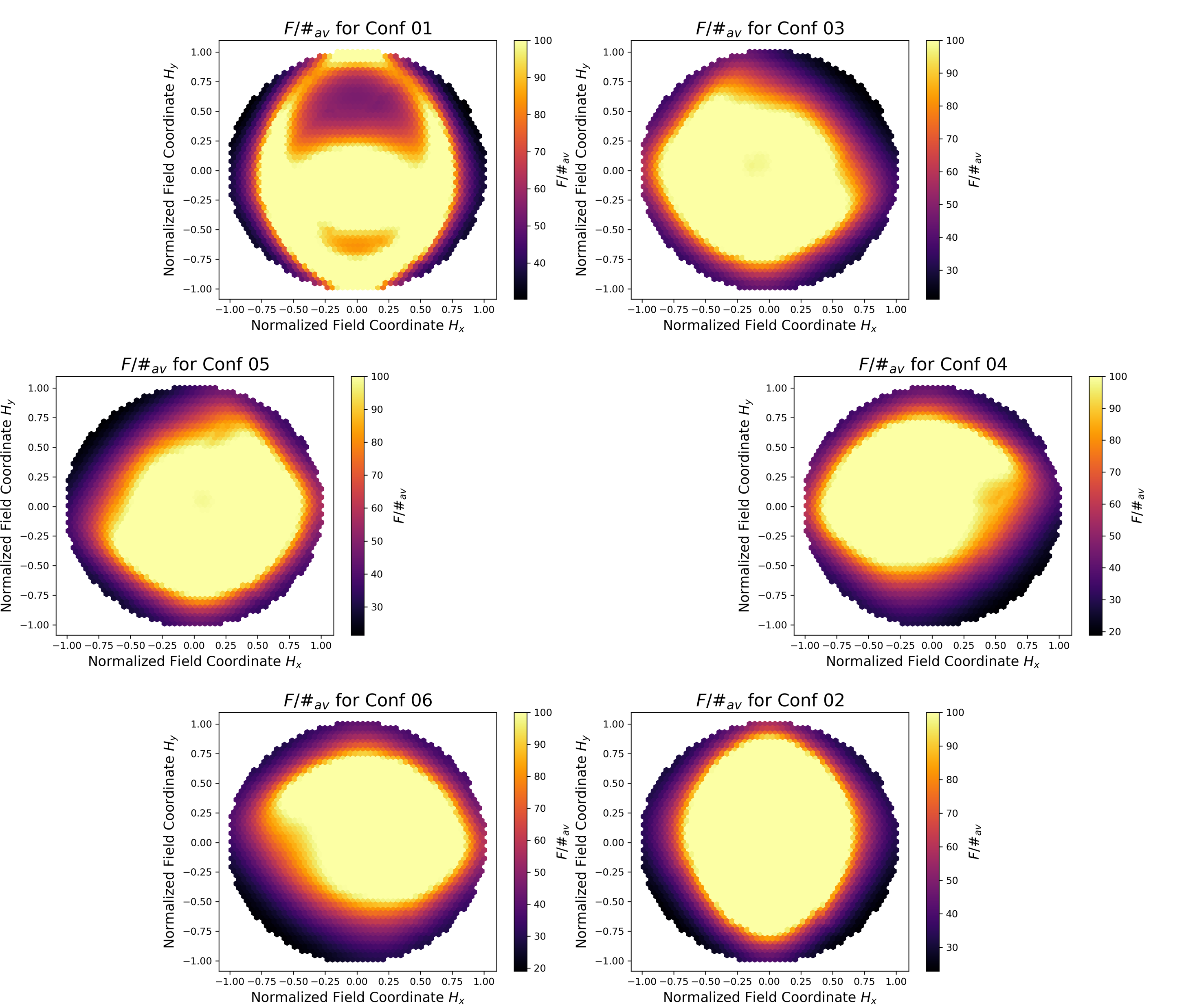}
    \caption{The F/\# at the Lyot stop for different fields across the focal plane at 60 degrees elevation. Any field exceeding the target value of 100 is shown in yellow. Because of the rotation used in the Zemax file for the receiver to simulate the elevation shift away from zenith (which produces the coordinate system shown in Fig.~\ref{fig:sixtubesreceiver}), each plot looks rotated by 30 degrees away from the symmetry axis of the telescope mirrors.}
    \label{fig:FNumLyot}
\end{figure}

To measure the collimation of the beam, we calculate the F/\# at the Lyot stop for many fields across our field of view. 
For each field, we trace the rays from that field to the extreme $+$x (far right), $-$x (far left), $+$y (top), and $-$y (bottom) locations on the Lyot stop, calculate the angle between the extreme rays in the x direction or the extreme rays in the y direction by taking a dot product, converting each angle to an F/\# using
\begin{equation}
    F/\# = \frac{1}{2 \tan(\theta)},
\end{equation}
and averaging the F/\# calculated from the extreme rays in the x direction and the F/\# calculated from the extreme rays in the y direction to get an estimate of the F/\# for that field. A higher F/\# indicates a more collimated beam.
In order to achieve the target FPI resolving power of 100, we aim for an F/\# of 100 for each field at the Lyot stop. The results for each configuration are shown in Fig.~\ref{fig:FNumLyot}. 
For locations 3-6, which also have the best image quality, around 40\% of the fields have an F/\# greater than 100 and 75\% of the fields have an F/\# greater than 50. The maximum F/\# for any field is around 250, while the minimum near the edges of the focal plane is near 20. 

\begin{figure}
    \centering
    \includegraphics[width=0.9\linewidth]{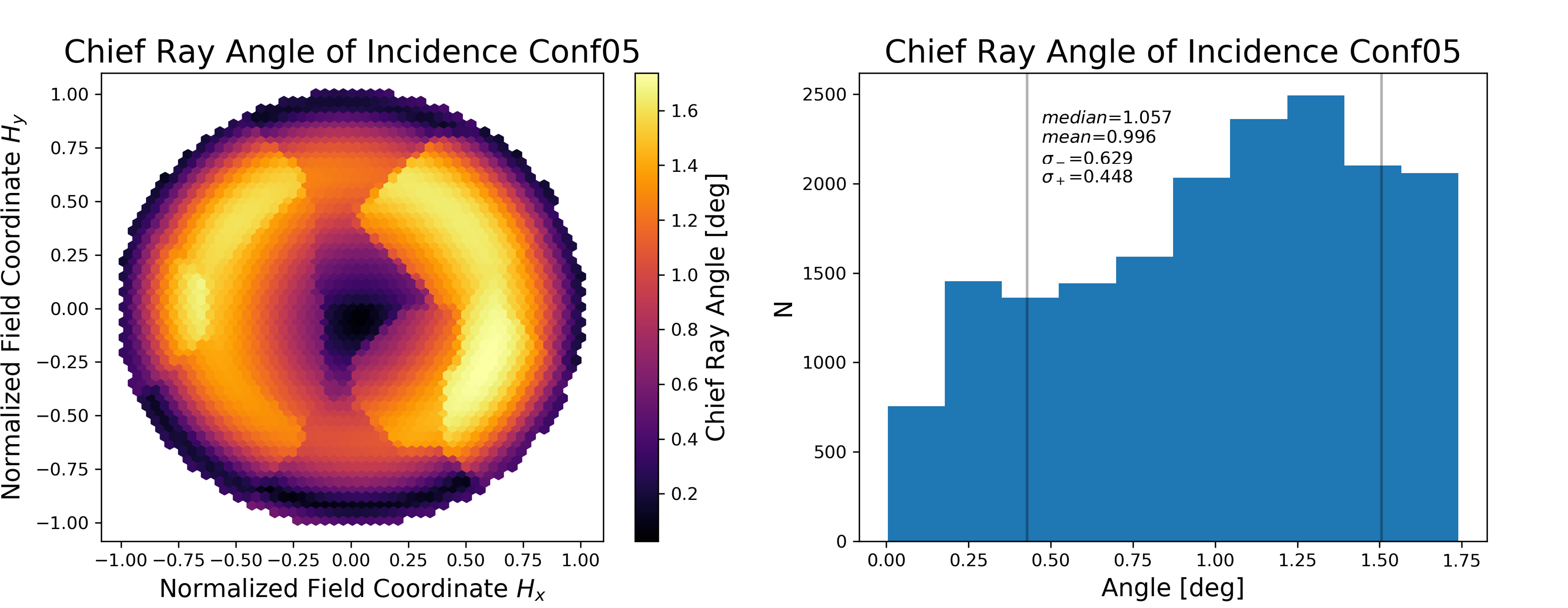}
    \caption{The chief ray angles across the focal plane must be smaller than two degrees to minimize reflections due to poor coupling to the detector feedhorns. On the left, the chief ray angles across the focal plane are plotted by location; the maximum chief ray angle is around 1.7. The histogram of these angles is shown on the right. All locations in the receiver show similar behavior.}
    \label{fig:chiefrayimage}
\end{figure}

The design also meets all other optical requirements for the cryostat and detector design outlined in Ref.~\citenum{Dicker2018}, including an underfilled primary mirror and a well-established Lyot stop that is the pupil-defining surface for all fields. The re-imaging optics set the focal plane size to be around 275 mm in diameter to maximize the area filled by the detector arrays. In order to ensure that the light couples well to the feedhorns on the detector array, the chief ray angles across the focal plane are restricted to two degrees or smaller, as shown in Fig.~\ref{fig:chiefrayimage} for a single configuration.

\section{BICONIC LENS COMPARISON}
\label{sec:biconic}
\begin{figure}
    \centering
    \includegraphics[width=0.8\linewidth]{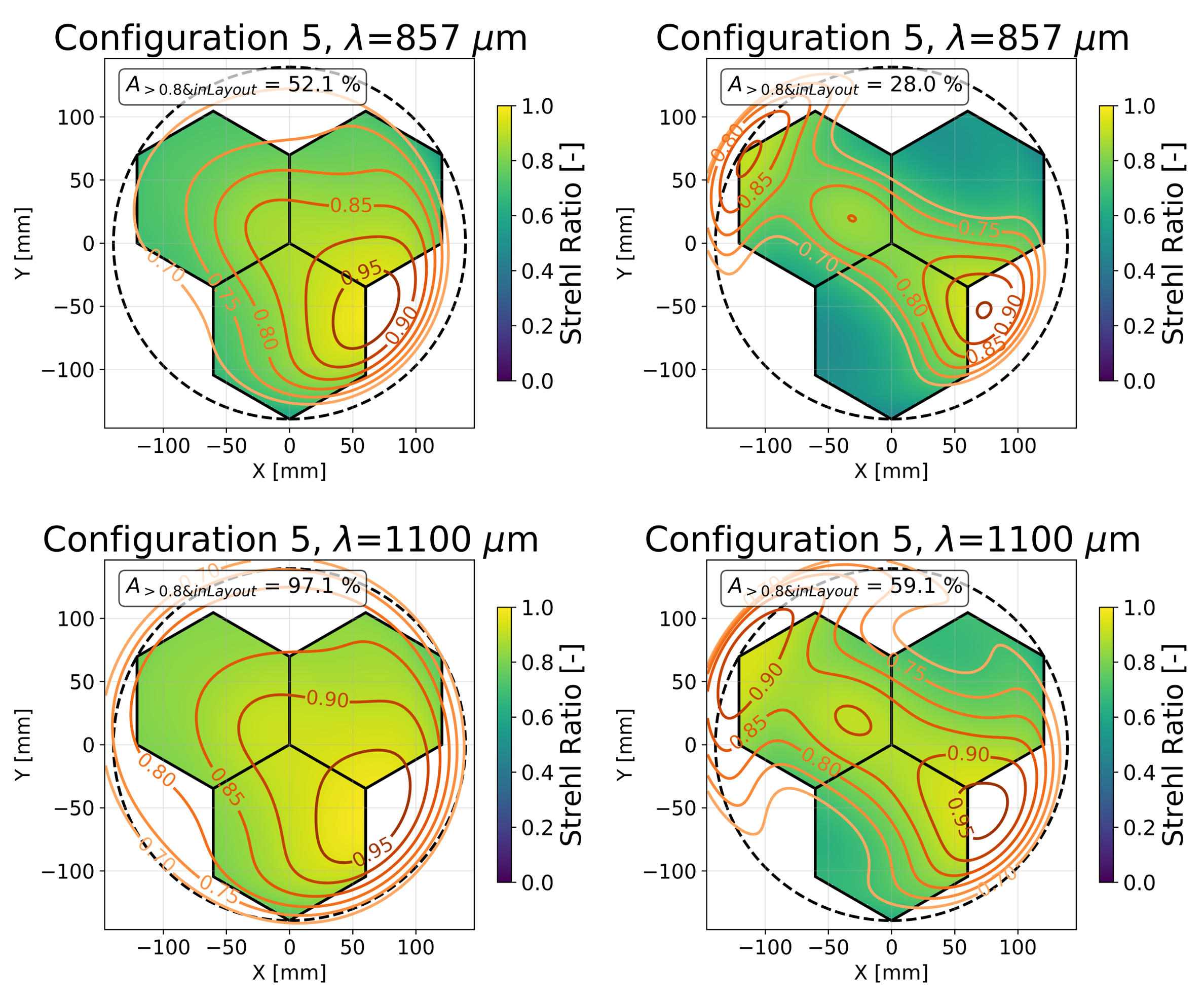}
    \caption{\textbf{Left:} The image quality at both detector array center wavelengths for a single configuration with lens three changed to a biconic lens. \textbf{Right:} The image quality at both detector array center wavelengths for a single configuration with four aspheric lenses. The fraction of the focal plane area filled by detector arrays with a Strehl ratio greater than 0.8 is 38.0\% and 24.1\% lower at 1.1 mm and 0.8 mm than the biconic design.}
    \label{fig:biconiccomp}
\end{figure}

Prior to changing lens three to a biconic lens, it was difficult to obtain a design with good image quality that also had a high F/\# at the Lyot stop for most fields in the field of view. The right column of Fig.~\ref{fig:biconiccomp} shows the best design obtained with good F/\# using aspheric lenses alone. The design shows a strong radial dependency with high Strehl quality along the axis running radially towards the center of the receiver and poor image quality off of that line.

The left column of Fig.~\ref{fig:biconiccomp} shows the same design after lens three is changed to a biconic lens and that design is optimized for image quality. Since lens three comes after the Lyot stop, this change did not alter the collimation of the beam at the stop, but it did significantly improve the image quality. The fraction of the focal plane area filled by detector arrays with a Strehl ratio greater than 0.8 is 37.1\% and 24.4\% higher at 1.1 mm and 0.8 mm, respectively, than the design that used aspheric lenses alone.

We also explored the possibility of changing lens four to a biconic lens, but such designs did not show a significant improvement of the image quality in limited testing. Future designs for similar FPI-based spectrometers in Prime-Cam or future receivers could continue to explore the conversion of another lens to a biconic lens. Further, more detailed studies of making lens four biconic could lead to an improvement of image quality. Another option would be to change lens two to a biconic lens, which would also impact the collimation of the beam at the stop.

\section{CONCLUSION}
\label{sec:conclusion}
In order to use EoR-Spec's FPI to make quality measurements of the 158 $\mu$m [CII] line to probe structure formation during the epoch of reionization, it is important to have a cold optical design that balances a highly collimated beam at the Lyot stop of the system with excellent image quality across the detector arrays on the focal plane. Other mechanical constraints on the sag of the lenses and the size of the focal plane must also be met. By utilizing three aspheric lenses and one biconic lens, this optical design ensures the majority of the focal plane is diffraction-limited near 1.1 mm and sufficient area is diffraction-limited near 0.8 mm for the single detector array centered at that frequency in EoR-Spec. It also produces a highly collimated beam for a significant portion of the field of view of the instrument across all six possible locations for EoR-Spec in Prime-Cam. In the coming year, machining of the lenses for the cold optics of EoR-Spec will begin in order to prepare the instrument for calibration and early science in 2024.

\acknowledgments % equivalent to \section*{ACKNOWLEDGMENTS} 

The CCAT-prime project, FYST and Prime-Cam instrument have been supported by generous contributions from the Fred M. Young, Jr. Charitable Trust, Cornell University, and the Canada Foundation for Innovation and the Provinces of Ontario, Alberta, and British Columbia. The construction of the FYST telescope was supported by the Gro{\ss}ger{\"a}te-Programm of the German Science Foundation (Deutsche Forschungsgemeinschaft, DFG) under grant INST 216/733-1 FUGG, as well as funding from Universit{\"a}t zu K{\"o}ln, Universit{\"a}t Bonn and the Max Planck Institut f{\"u}r Astrophysik, Garching.
The construction of EoR-Spec is supported by NSF grant AST-2009767. ZBH acknowledges support from a NASA Space Technology Graduate Research Opportunities Award. MDN acknowledges support from NSF grant AST-2117631. SKC acknowledges support from NSF award AST-2001866.

% References
\bibliography{main} % bibliography data in main.bib
\bibliographystyle{spiebib} % makes bibtex use spiebib.bst

\end{document}